\documentclass[11pt]{article}
\usepackage{titling}
\usepackage[OT1]{fontenc}
\usepackage{amsmath}
\usepackage{mathrsfs}
\usepackage{amsfonts}
\usepackage{amssymb}
\usepackage{diagbox}
\usepackage{footnote}
\usepackage{graphicx}
\usepackage{float}
\usepackage{epstopdf}
\usepackage[margin=1.25in]{geometry}
\usepackage{setspace}
\usepackage{tikz}
\usepackage{pgfplots}
\pgfplotsset{compat=1.18}
\usepackage[round]{natbib}
\usepackage{multirow}
\usepackage[toc,page]{appendix}
\usepackage{hyperref}
\usepackage{subcaption}
\usepackage{enumitem}
\usepackage{booktabs}
\usepackage{chapterbib}
\usepackage{dsfont}
\usepackage{algorithm}
\usepackage{longtable}
\usepackage{caption}
\usepackage[flushleft]{threeparttable}
\usepackage[flushleft]{threeparttablex}
\usepackage{pdflscape}
\DeclareMathOperator*{\argmax}{arg\,max}
\providecommand{\U}[1]{\protect\rule{.1in}{.1in}}
\providecommand{\keywords}[1]{\textsc{Keywords}: #1}

\hypersetup{
colorlinks=true,
linkcolor=blue,
citecolor=blue,
filecolor=magenta,
urlcolor=blue,
breaklinks=true,
}

\DeclareMathOperator{\median}{Med}

\begin{document}
\title{Bootstrap Inference on Partially Linear Binary Choice Model\thanks{This work was supported by the National Natural Science Foundation of China [Grant Numbers 72103004, 72103101].}}
\author{Wenzheng Gao\\ School of Economics \\ Nankai University \\ wenzhenggao@nankai.edu.cn 
\and Zhenting Sun\\ China Center for Economic Research \\ National School of Development \\ Peking University \\ zhentingsun@nsd.pku.edu.cn 
}
\maketitle

\begin{abstract}
The partially linear binary choice model
can be used for estimating structural equations where nonlinearity may appear due to diminishing
marginal returns, different life cycle regimes, or hectic physical phenomena. The inference procedure for this model based on the analytic asymptotic approximation could be unreliable in finite samples if the sample size is not sufficiently large. This paper proposes a bootstrap inference approach for the model. Monte Carlo simulations show that the proposed inference method performs well in finite samples compared to the procedure based on the asymptotic approximation. 

\bigskip
\keywords{Partially linear binary choice model, bootstrap inference, good finite sample properties}  
\end{abstract}
\bigskip

\newpage
\section{Introduction}
In this paper, we propose a bootstrap inference procedure for the partially linear binary choice model studied by \cite{krief2014integrated}. 
As introduced in \citet{krief2014integrated}, this model is useful for estimating structural equations where nonlinearity is suspected, possibly arising from diminishing marginal returns, different life cycle regimes,
or hectic physical phenomena. This model may also encompass the endogenous binary response model with a median control
function restriction \citep{blundell2004endogeneity} and the binary response model with a partially nonadditive error. \citet{krief2014integrated} proposes a two-stage smoothed maximum score (SMS) estimator for the coefficient vector in the model and provides the asymptotic distribution for the estimator.

For inference of the model, \citet{horowitz2002bootstrap} and \citet{cao2021smoothed} show that the hypothesis testing results in finite samples based on the SMS estimator using analytic asymptotic approximation may not be reliable if the sample size is not sufficiently large. We follow the idea of  \citet{horowitz2002bootstrap} and \citet{cao2021smoothed} and propose a bootstrap inference approach for the partially linear binary
choice model. Our simulation results show that the proposed method can significantly address the over rejection issue caused by using the analytic asymptotic approximation, while maintaining a relatively high level of power. 

\section{Setup and Inference Procedure}
We consider the partially linear binary choice model in \citet{krief2014integrated}:
\begin{equation}\label{eqn: model}
Y=\mathds{1}(X^T\beta_X+\phi(V)+\epsilon\geq 0),
\end{equation}
where $Y$ is a binary dependent variable, $X$ is a $d\times 1$ vector of independent variables, $\beta_X$ is the corresponding vector of unknown coefficients, $V$ is a continuously distributed univariate, $\phi $ is an unknown real-valued function, and $\epsilon$ is an unobservable error term.

For the purpose of identification, we follow \citet{krief2014integrated} and assume that the first component of $X$, denoted by $X_{1}$, conditional on both the remaining components of $X$, denoted by $\widetilde{X}$, and $V$ admit a distribution function absolutely continuous with respect to the Lebesgue measure almost surely.
Following \cite{horowitz1992smoothed}, \citet{krief2014integrated}, and \cite{chen2015binary}, we assume $\beta_{1}=1$ for scale normalization, where $\beta_{1}$ is the coefficient of $X_{1}$. Let $\beta$ denote the coefficients for $\Tilde{X}$.

Under the assumption that the conditional median of $\epsilon$ given $X$ and $V$ is zero, i.e., $\median(\epsilon|X,V)=0$, \cite{krief2014integrated} proposed the integrated kernel-weighted smoothed maximum score (IKWSMS) estimator for $\beta$ in a two-stage procedure. 
Let $ G(\cdot) $ be a continuous function that corresponds to the integral of a $r$-th order kernel function for some integer $r\geq 4$.\footnote{See, for example, the definition of a $r$-th order kernel function in \citet[p.~33]{li2006nonparametric}.} We assume that  $\sup_{t\in\mathbb{R}} |G(t)|<M $ for some $ M<\infty $, $ \lim_{t\rightarrow-\infty}G(t)=0 $, and $ \lim_{t\rightarrow\infty}G(t)=1 $. Let $K(\cdot)$ be another kernel function.
Suppose we observe an i.i.d.\ sample $\{(Y_i,X_i,V_i)\}_{i=1}^n$ from the distribution of $(Y,X,V)$. 

In the first stage, we fix a $v$ in the support $\mathcal{V}$ of $V$ and estimate $\beta$ for $v$ by
\begin{equation}\label{eqn: 1st stage}
\beta_{n}(v)=e_{d}\argmax_{\theta\in\Theta}\dfrac{1}{nh_{v}}\sum_{i=1}^{n}(2Y_{i}-1)G\left(\dfrac{X_{i1}+W_{i}^T\theta}{h}\right)K\left(\dfrac{V_{i}-v}{h_{v}}\right),
\end{equation}
where $W_i^T=(1,\widetilde{X}_i^T)$, $e_{d}=[O,I_{d-1}]$ is a $(d-1)\times d$ matrix with the first column being a zero vector, $I_{d-1}$ is the $(d-1)\times (d-1)$ identity matrix, $\Theta\subset \mathbb{R}^d$ is some compact set, and $h$ and $h_{v}$ are the smoothing parameters that converge to $0$ as $n\rightarrow\infty$. In this way, we obtain the estimator $\beta_{n}(v)$ for every $v\in\mathcal{V}$.
In the second stage, the IKWSMS estimator, denoted by $\widehat{\beta}$, is defined as
\begin{equation}\label{eqn: 2nd stage}
    \widehat{\beta}=\int_{\mathcal{V}}\beta_{n}(v)\tau(v)dv,
\end{equation}
where $\tau(\cdot)$ is a known weighting function on the real line with compact support that satisfies
\begin{equation*}
   \int_{\mathcal{V}}\tau(v)dv=1.
\end{equation*}

For each $v$, define $L(v)=X^T\beta_X+\phi(v)$. 
Let $F_{\epsilon|W,L(v),V}$, $f_{V|W,L(v)}$, and $f_{L(v)|W}$ be the conditional cumulative distribution function (CDF) of $\epsilon$ given $(W,L(v),V)$, the conditional probability density funciton (PDF) of $V$ given $(W,L(v))$, and the conditional PDF of $L(v)$ given $W$, respectively. Then we define the random elements
\begin{equation*}
    T_{W}(l,v)=[1-2F_{\epsilon|W,L(v),V}(-l|W,l,v)]f_{V|W,L(v)}(v|W,l)f_{L(v)|W}(l|W)
\end{equation*}
and $T_{W}^{(1)}(l,v)=\partial T_{W}(l,v)/\partial l$ whenever these quantities exist. Also, we define $Z=X^T\beta_X+\phi(V)$. Then we let $F_{\epsilon|Z,W,V}$ and $f_{Z|W,V}$ be the conditional CDF of $\epsilon$ given $(Z,W,V)$ and the condition PDF of $Z$ given $(W,V)$, respectively. We define the random elements
\begin{equation*}
[1-2F_{\epsilon|Z,W,V}(-z|z,W,V)]^{(j)}=\dfrac{\partial^{j}[1-2F_{\epsilon|Z,W,V}(-z|z,W,V)]}{\partial z^{j}}
\end{equation*}
and
\begin{equation*}
[f_{Z|W,V}(z|W,V)]^{(j)}=\dfrac{\partial^{j}f_{Z|W,V}(z|W,V)}{\partial z^{j}}
\end{equation*}
for $j=1,\ldots,r$, provided these derivatives exist. Furthermore, for every $v$ such that all the relevant elements exist, we define
\begin{equation*}
Q(v)=\tau(v)e_{d}\left\{\mathbb{E}[WW^TT_{W}^{(1)}(0,v)]\right\}^{-1},
\end{equation*}
\begin{align*}
B=&\,\sum_{s=0}^{r}\dfrac{1}{s!(r-s)!}\mathbb{E}\left\{Q(V)W[1-2F_{\epsilon|Z,W,V}(0|0,W,V)]^{(s)}[f_{Z|W,V}(0|W,V)]^{(r-s)}\right\}\\
&\,\cdot\left[\int t^{r}G'(t)dt\right],
\end{align*}
and
\begin{equation*}
\Omega=\left\{\int \left[G'(t)\right]^{2}dt\right\}\mathbb{E}\left[f_{Z|W,V}(0|W,V)Q(V)WW^TQ(V)^T\right].
\end{equation*}
Under certain conditions, \citet[Theorem 1]{krief2014integrated} showed that 
\begin{equation}
\sqrt{nh}\left(\widehat{\beta}-\beta\right)\xrightarrow{d}\mathcal{N}(\lambda B, \Omega),
\end{equation}
where $\lambda=\lim_{n\rightarrow\infty}\sqrt{nh}h^{r}<\infty$.

It has been shown in \cite{horowitz2002bootstrap} and \cite{cao2021smoothed} that hypothesis testing of the SMS estimator based on asymptotic critical values may not have good finite sample performances. This is because in nonparametric and semiparametric estimations, large samples may be required to achieve reasonable agreement between the finite sample properties and the results of asymptotic theory. We next suggest using a bootstrap method for inference that could provide refinements to the empirical size of the test in finite samples.

For inference, we first consider testing the hypothesis 
\begin{align}\label{eq.hypothesis t test}
H_{0}: \beta_{j}=\beta_{0j} \text{ against } H_{1}:\beta_{j}\neq\beta_{0j},    
\end{align}
where $\beta_{j}$ denotes the $j$th component of $\beta$ and $\beta_{0j}$ is some prespecified value. To avoid computing the complicated bias term when conducting the test, we follow \cite{horowitz2002bootstrap} and use the undersmoothed bandwidth $h$ such that $h$ converges to zero sufficiently fast so that $\lambda=0$. Thus, the test statistic for the $H_0$ in \eqref{eq.hypothesis t test} can be constructed as
\begin{equation}\label{eqn: t test}
 t_{n}=\dfrac{(nh)^{1/2}\left(\widehat{\beta}_{j}-\beta_{0j}\right)}{\widehat{\Omega}_{j}^{1/2}},
\end{equation}
where $\widehat{\beta}_j$ denotes the $j$th component of $\widehat{\beta}$ and $\widehat{\Omega}_{j}$ is the $(j,j)$ component of $\widehat{\Omega}$, a consistent estimator of $\Omega$. Based on Lemma 2 and the proof of Theorem 1 in \citet{krief2014integrated}, such an estimator may be constructed as
\begin{equation*}\label{eq.estimator of Omega}
\widehat{\Omega}=\frac{1}{nh}\sum_{j=1}^{n}\widehat{Q}(V_{j})W_{j}W_{j}^{T}\widehat{Q}(V_{j})^{T}\left[G'\left(\dfrac{X_{j1}+W_{j}^T\widehat\theta(V_{j})}{h}\right)\right]^{2},
\end{equation*}
where 
\begin{equation}\label{eq.1st stage at V_j}
\widehat\theta(V_{j})=\argmax_{\theta\in\Theta}\dfrac{1}{nh_{v}}\sum_{i=1}^{n}(2Y_{i}-1)G\left(\dfrac{X_{i1}+W_{i}^T\theta}{h}\right)K\left(\dfrac{V_{i}-V_{j}}{h_{v}}\right),
\end{equation}
\begin{equation*}
\widehat{Q}(V_{j})=\tau(V_{j})e_{d}[-H(V_{j})]^{-1},
\end{equation*}
and $H(V_{j})$ is the Hessian matrix of the objective function of $\theta$ in (\ref{eq.1st stage at V_j}) evaluated at $\widehat{\theta}(V_j)$.
Algorithm \ref{proc:bootstrap t} illustrates the bootstrap procedure for the test.

\begin{algorithm}\caption{Bootstrap $t$ Test }\label{proc:bootstrap t}	
	\begin{enumerate}[label=(\arabic*)]
    
\item Draw a bootstrap sample $\{(Y_{i}^{*},X_{i}^{*},V_{i}^{*})\}_{i=1}^{n}$ independently from the sample $\{(Y_{i},X_{i},V_{i})\}_{i=1}^{n}$ with replacement.

\item Using the bootstrap sample, compute the bootstrap version of the IKWSMS estimator $\widehat{\beta}^{*}$ and the bootstrap estimator $\widehat\Omega^*$ for $\Omega$, and obtain the bootstrap test statistic
\begin{equation*}
t^{*}_{n}=\dfrac{(nh)^{1/2}\left(\widehat{\beta}^{*}_{j}-\widehat{\beta}_{j}\right)}{\widehat{\Omega}_{j}^{*1/2}}.
\end{equation*}

\item Repeat steps (1) and (2) many times.
For a given nominal significance level $\alpha$, we construct the bootstrap critical value
		$\widehat{c}_{1-\alpha}$ by
		\begin{align*}
			\widehat{c}_{1-\alpha}=\inf\left\{  c:\mathbb{P}\left(  |t_n^*|  \leq c\big|\{(Y_{i},D_{i},Z_{i})\}_{i=1}^{n}\right)  \geq
			1-\alpha\right\}  .
		\end{align*}
		In practice, we may approximate $\widehat{c}_{1-\alpha}$ by the $1-\alpha$ quantile of the independently generated bootstrap statistics.

\item The decision rule of the test is reject $H_0$ if $|t_n|>\widehat c_{1-\alpha}$.

\end{enumerate}
\end{algorithm}

We are also interested in a more general hypothesis 
\begin{align}\label{eq.hypothesis W test}
H_0: \mathcal{F}(\beta)=0 \text{ against } H_1: \mathcal{F}(\beta)\neq0,    
\end{align}
where $\mathcal{F}$ can be a vector valued function with continuous first derivatives. One special case is that $\mathcal{F}(b)=Rb$ with $b\in\mathbb{R}^{d-1}$ for some suitable matrix $R$ which incorporates the case where $H_{0}: \beta_{j}=\beta_{0j}$. Let $\mathcal{F}'(b)=\partial\mathcal{F}(b)/\partial b^T$ for all suitable $b\in\mathbb{R}^{d-1}$. Then we define the test statistic for the $H_0$ in \eqref{eq.hypothesis W test} as
\begin{align}\label{eq.test stat W}
    W_n=(nh)\mathcal{F}(\widehat{\beta})^T\left\{ \mathcal{F}'(\widehat{\beta})\widehat{\Omega}\mathcal{F}'(\widehat{\beta})^T\right\}^{-1}\mathcal{F}(\widehat{\beta}).
\end{align}
Algorithm \ref{proc:bootstrap W} illustrates the bootstrap procedure for the test. 

\begin{algorithm}\caption{Bootstrap Wald Test }\label{proc:bootstrap W}	
	\begin{enumerate}[label=(\arabic*)]
    
\item Draw a bootstrap sample $\{(Y_{i}^{*},X_{i}^{*},V_{i}^{*})\}_{i=1}^{n}$ independently from the sample $\{(Y_{i},X_{i},V_{i})\}_{i=1}^{n}$ with replacement.

\item Using the bootstrap sample, compute the bootstrap version of the IKWSMS estimator $\widehat{\beta}^{*}$ and the bootstrap estimator $\widehat\Omega^*$ for $\Omega$, and obtain the bootstrap test statistic
\begin{align}\label{eq.bootstrap test stat W}
    W_n^*=(nh)(\mathcal{F}(\widehat{\beta}^*)-\mathcal{F}(\widehat{\beta}))^T\left\{ \mathcal{F}'(\widehat{\beta}^*)\widehat{\Omega}^*\mathcal{F}'(\widehat{\beta}^*)^T\right\}^{-1}(\mathcal{F}(\widehat{\beta}^*)-\mathcal{F}(\widehat{\beta})).
\end{align}

\item Repeat steps (1) and (2) many times.
For a given nominal significance level $\alpha$, we construct the bootstrap critical value
		$\widehat{c}_{1-\alpha}$ by
		\begin{align*}
			\widehat{c}_{1-\alpha}=\inf\left\{  c:\mathbb{P}\left(  W_n^*  \leq c\big|\{(Y_{i},D_{i},Z_{i})\}_{i=1}^{n}\right)  \geq
			1-\alpha\right\}  .
		\end{align*}
		In practice, we may approximate $\widehat{c}_{1-\alpha}$ by the $1-\alpha$ quantile of the independently generated bootstrap statistics.

\item The decision rule of the test is reject $H_0$ if $W_n>\widehat c_{1-\alpha}$.

\end{enumerate}
\end{algorithm}

The asymptotic properties of the bootstrap tests in Algorithms \ref{proc:bootstrap t} and \ref{proc:bootstrap W} may be proved analogously to Theorems 4.3 in \citet{horowitz2002bootstrap}. To avoid theoretical complications, we omit the proofs in the paper and show the finite sample properties of the tests in Monte Carlo simulations.

\section{Monte Carlo Simulations}

In this section, we present the Monte Carlo investigation of the finite sample performance of the proposed method. We design simulations based on those in \cite{horowitz2002bootstrap} and \cite{krief2014integrated}. Specifically, we consider estimating $\beta$ in the following model:
\begin{equation*}
Y=\mathds{1}(X_{1}+\beta X_{2}+\phi(V)+\epsilon\geq 0),
\end{equation*}
where $\beta=1$, $\phi(v)=\cos(2\pi v)$ for every $v$, $V=\Phi(\xi)$ with $\Phi(\cdot)$ being the cumulative distribution function of the standard normal random variable, and $(X_{1},X_{2},\xi)$ is a standard normal triplet with correlation coefficient $0.2$. We consider five distributions for the error $\epsilon$ as follows:
\begin{enumerate}
    \item \textbf{UN}: $\epsilon\sim \mathrm{Unif}(-\sqrt{3},\sqrt{3})$
    \item \textbf{NR}: $\epsilon\sim \mathcal{N}(0,1)$
    \item \textbf{T3}: $\epsilon\sim t_3$ (Student's $ t $ distribution with 3 degrees of freedom)

    \item \textbf{LG}: $\epsilon\sim \mathrm{Logistic}(0,\pi/3)$ (logistic distribution)

    \item \textbf{HE}: $\epsilon=(1+X_{1}^{2}+X_{2}^{2}+V^{2})e$, where $e\sim \mathrm{Logistic}(0,\pi/3)$ and $e\perp (X_1,X_2,V)$
\end{enumerate}

In all designs, $\epsilon$ is normalized with variance $1$. Every Monte Carlo experiment consists of $500$ replications. The warp-speed method of \cite{giacomini2013warp} is employed
to expedite the simulations. Similar to \cite{krief2014integrated}, the smoothing function for the indicator function is 
\begin{equation*}
G(t)=\left[0.5+\frac{105}{64}\left(t-\frac{5}{3}t^{3}+\frac{7}{5}t^{5}-\frac{3}{7}t^{7}\right)\right]\mathds{1}(|t|\leq 1)+\mathds{1}(t>1).
\end{equation*}
The derivative of $G$ is a kernel function with order $r=4$ \citep{muller1984smooth}. The first-stage kernel function $K$ is
\begin{equation*}
K(t)=\frac{1}{48}\left(105-105t^{2}+21t^{4}-t^{6}\right)\Phi'(t),
\end{equation*}
which is of order $8$ \citep{pagan1999nonparametric}. The bandwidth selection method follows \cite{krief2014integrated} (Silverman-like rule of thumb in \citet{silverman1986density}). Specifically, for every $v$, we let $h=1.22 R_{L}n^{-1/(2r+1)}$ and $h_{v}=0.8R_{V}n^{-3/25}$, where  $R_{L}$ and $R_{V}$ denote the interquartile ranges between the first and third quartiles of $\{X_{i1}+(1,X_{i2})\widehat{\theta}(v)\}_{i=1}^{n}$ and $\{V_{i}\}_{i=1}^{n}$, respectively, and $\widehat\theta(v)$ is obtained with $(h_0,h_{v0})=(n^{-1/(2r+1)},n^{-3/25})$ by 
\begin{equation}\label{eqn: 1st stage 2}
\widehat\theta(v)=\argmax_{\theta\in\Theta}\dfrac{1}{nh_{v0}}\sum_{i=1}^{n}(2Y_{i}-1)G\left(\dfrac{X_{i1}+(1,X_{i2})\theta}{h_0}\right)K\left(\dfrac{V_{i}-v}{h_{v0}}\right).
\end{equation}
The weighting function $\tau(v)=1.25\cdot\mathds{1}(0.1\leq v \leq 0.9)$.

First, we run simulations to show the size control of the test. Both the asymptotic and bootstrap critical values are used to compute the empirical size of the $t$ test for $H_{0}: \beta=1$. The sample size $n$ is set to $1000$. The results of the experiments are shown in Table \ref{tb: size}. In contrast to the results obtained from using the asymptotic critical values, the empirical sizes obtained from using the bootstrap critical values are well controlled. Furthermore, the empirical sizes obtained from using the bootstrap critical values show little sensitivity for undersmoothed bandwidths $0.75h$ and $0.5h$. They are all well controlled for undersmoothed bandwidths.

\begin{table}[htbp]
	\caption{Empirical Sizes of $t$ Tests Using Asymptotic and Bootstrap Critical Values}\label{tb: size}
 \centering
	\begin{threeparttable}
	\begin{tabular}{lccccccc}
		\hline
		\hline
		 \multicolumn{1}{l}{}	&   & \multicolumn{2}{c}{$h$} & \multicolumn{2}{c}{$0.75h$}& \multicolumn{2}{c}{$0.5h$}\\ \cmidrule(lr){3-4} \cmidrule(lr){5-6} \cmidrule(lr){7-8}
		 Model  & $\alpha$ & ACV&BCV& ACV&BCV & ACV&BCV\\
		 \hline
		                         &  0.10  &   0.178    & 0.094  & 0.192 & 0.088 & 0.260 & 0.056\\
		             UN          &  0.05  &  0.114   & 0.036 & 0.150 & 0.022 & 0.212 & 0.024 \\
		                       &  0.01 &  0.044     & 0.004 & 0.080 & 0.002 & 0.152 & 0.008\\
          \hline
		                          &  0.10  &   0.172   & 0.150 & 0.190 & 0.098 & 0.254 & 0.094\\
		               NR        &  0.05  &  0.116     & 0.048 & 0.152 & 0.034 & 0.210 & 0.034\\
		                          &  0.01 &  0.052    & 0.008 & 0.090 & 0.000 & 0.142 & 0.006\\
		 \hline
		                         &  0.10  &  0.172    & 0.102 & 0.162 & 0.094 & 0.206 & 0.088\\
		           T3          &  0.05  &  0.104     &0.048 & 0.120 & 0.020 & 0.170 & 0.022\\
		                         &  0.01 &  0.036     &0.006 & 0.052 & 0.004 & 0.088 & 0.000\\
		 \hline  
		                         &  0.10  &  0.196      & 0.140  & 0.172 & 0.096 & 0.244 & 0.068\\
		         LG              &  0.05  &   0.126      &0.062 & 0.136 & 0.046 & 0.196 & 0.038\\
		                         &  0.01 &   0.050    &0.002  & 0.082 & 0.006 & 0.118 & 0.004\\
		 \hline
		                         &  0.10  &   0.224   &0.120 & 0.188 & 0.086 & 0.272 & 0.106\\
		          HE            &  0.05  &  0.150   &0.040 & 0.140 & 0.038 & 0.228 & 0.048\\
		                         &  0.01 &  0.074    &0.010 & 0.064 & 0.000 & 0.132 & 0.002\\
		 \hline      
\multicolumn{4}{l}{}
	\end{tabular}
 \begin{tablenotes}[para,flushleft]
\footnotesize{\textit{Note}: This table shows the empirical sizes of the $t$ tests using both the asymptotic critical values (ACVs) and bootstrap critical values (BCVs) for the null hypothesis $\beta=1$ with various nominal levels. The sample size is $n=1000$.}
\end{tablenotes}
\end{threeparttable}
\end{table}

We also investigate the empirical power of the test. The sample sizes we consider are $n\in\{250,500,1000\}$. Table \ref{tb: power} reports the rejection rates of the $t$ test for the false hypothesis $H_{0}: \beta=0$. The empirical powers obtained from using the asymptotic critical values exceed those obtained from using the bootstrap critical values. This is natural because of the upward size distortions arising from using the asymptotic critical values. But we note that the differences are small.
Furthermore, the empirical power of the proposed bootstrap test increases as the sample size increases, which demonstrates the good finite sample power property of the proposed test.

\begin{table}[htbp!]
	\caption{Empirical Powers of $t$ Tests Using Asymptotic and Bootstrap Critical Values}\label{tb: power}
	\centering
	\begin{threeparttable}
\begin{tabular}{lccccccc}
		\hline
		\hline
	\multicolumn{1}{l}{}	&   & \multicolumn{2}{c}{$\alpha=0.1$} & \multicolumn{2}{c}{$\alpha=0.05$}& \multicolumn{2}{c}{$\alpha=0.01$}\\ \cmidrule(lr){3-4} \cmidrule(lr){5-6} \cmidrule(lr){7-8}
		 Model  & $n$ & ACV&BCV& ACV&BCV & ACV&BCV\\
\hline
      &  250  &   0.696    & 0.578 &   0.664    & 0.450 &    0.622    & 0.202  \\
   UN   & 500 &  0.680     & 0.664 & 0.656     & 0.604 &  0.612     & 0.498\\
      & 1000 &  0.732     & 0.686 &  0.690     & 0.656 &  0.656     & 0.564\\    
\hline
      &  250  &   0.758   & 0.714  &   0.724   & 0.630 & 0.686   & 0.400 \\
  NR  & 500   &  0.718    & 0.692  &  0.698    & 0.656 & 0.666   & 0.500\\
      &  1000 &  0.792    & 0.788  &  0.776    & 0.736 &  0.738  & 0.664\\
\hline
      &  250  &  0.800   & 0.770   &  0.778   & 0.714  &0.738    & 0.374\\
T3    &  500  &  0.818   & 0.796   &  0.796   &0.768   & 0.768    &0.692\\
      &  1000 &  0.876   &0.870    &  0.870   &0.848   &0.844     &0.800\\
\hline
      &  250  &  0.720    & 0.618  & 0.700   & 0.574  &0.652      & 0.268\\
 LG   &  500  &   0.756   & 0.744 &  0.732   &0.700  &  0.708   &0.590\\
      &  1000 &   0.826   & 0.814 &  0.808  &0.782   &   0.780    &0.716\\
\hline
     &  250  &   0.844   &0.780  &   0.824   &0.690 &   0.774   &0.486\\
 HE  &  500  &  0.840   &0.826   &  0.822   &0.788  &  0.784   &0.706\\
     &  1000 &  0.864    &0.852  &  0.856    &0.824 &  0.838    &0.792\\
\hline
\end{tabular}
 \begin{tablenotes}[para,flushleft]
\footnotesize{\textit{Note}: This table shows the empirical powers of the $t$ tests using both the asymptotic critical values (ACVs) and bootstrap critical values (BCVs) for the null hypothesis $\beta=0$ with different sample sizes and nominal levels.}
\end{tablenotes}
\end{threeparttable}
\end{table}


\bibliographystyle{chicago}
\bibliography{Ref}

\begin{thebibliography}{}

\bibitem[\protect\citeauthoryear{Blundell and Powell}{Blundell and
  Powell}{2004}]{blundell2004endogeneity}
Blundell, R.~W. and J.~L. Powell (2004).
\newblock Endogeneity in semiparametric binary response models.
\newblock {\em The Review of Economic Studies\/}~{\em 71\/}(3), 655--679.

\bibitem[\protect\citeauthoryear{Cao, Chen, Gao, and Hsiao}{Cao
  et~al.}{2021}]{cao2021smoothed}
Cao, X., X.~Chen, W.~Gao, and C.~Hsiao (2021).
\newblock Smoothed maximum score estimation with nonparametrically generated
  covariates.
\newblock {\em Econometric Reviews\/}~{\em 40\/}(8), 796--813.

\bibitem[\protect\citeauthoryear{Chen and Zhang}{Chen and
  Zhang}{2015}]{chen2015binary}
Chen, S. and H.~Zhang (2015).
\newblock Binary quantile regression with local polynomial smoothing.
\newblock {\em Journal of Econometrics\/}~{\em 189\/}(1), 24--40.

\bibitem[\protect\citeauthoryear{Giacomini, Politis, and White}{Giacomini
  et~al.}{2013}]{giacomini2013warp}
Giacomini, R., D.~N. Politis, and H.~White (2013).
\newblock A warp-speed method for conducting monte carlo experiments involving
  bootstrap estimators.
\newblock {\em Econometric Theory\/}~{\em 29\/}(3), 567--589.

\bibitem[\protect\citeauthoryear{Horowitz}{Horowitz}{1992}]{horowitz1992smoothed}
Horowitz, J.~L. (1992).
\newblock A smoothed maximum score estimator for the binary response model.
\newblock {\em Econometrica\/}~{\em 60\/}(3), 505--531.

\bibitem[\protect\citeauthoryear{Horowitz}{Horowitz}{2002}]{horowitz2002bootstrap}
Horowitz, J.~L. (2002).
\newblock Bootstrap critical values for tests based on the smoothed maximum
  score estimator.
\newblock {\em Journal of Econometrics\/}~{\em 111\/}(2), 141--167.

\bibitem[\protect\citeauthoryear{Krief}{Krief}{2014}]{krief2014integrated}
Krief, J.~M. (2014).
\newblock An integrated kernel-weighted smoothed maximum score estimator for
  the partially linear binary response model.
\newblock {\em Econometric Theory\/}~{\em 30\/}(3), 647--675.

\bibitem[\protect\citeauthoryear{Li and Racine}{Li and
  Racine}{2006}]{li2006nonparametric}
Li, Q. and J.~S. Racine (2006).
\newblock {\em Nonparametric Econometrics: Theory and Practice}.
\newblock Princeton University Press.

\bibitem[\protect\citeauthoryear{M\"{u}ller}{M\"{u}ller}{1984}]{muller1984smooth}
M\"{u}ller, H. (1984).
\newblock Smooth optimum kernel estimators of densities, regression curves and
  modes.
\newblock {\em The Annals of Statistics\/}~{\em 12\/}(2), 766--774.

\bibitem[\protect\citeauthoryear{Pagan and Ullah}{Pagan and
  Ullah}{1999}]{pagan1999nonparametric}
Pagan, A. and A.~Ullah (1999).
\newblock {\em Nonparametric Econometrics}.
\newblock Cambridge University Press.

\bibitem[\protect\citeauthoryear{Silverman}{Silverman}{1986}]{silverman1986density}
Silverman, B.~W. (1986).
\newblock {\em Density Estimation}.
\newblock Chapman and Hall.

\end{thebibliography}

\end{document}